# Precision spectroscopy of non-thermal molecular plasmas using mid-infrared optical frequency comb Fourier transform spectroscopy

Ibrahim Sadiek,[1,*] Alexander Puth,[1] Grzegorz Kowzan,[2] Akiko Nishiyama,[2] Sarah-Johanna Klose,[1] Jürgen Röpcke,[1] Norbert Lang,[1] Piotr Masłowski,[2] Jean-Pierre H. van Helden[1]

[1]Leibniz Institute for Plasma Science and Technology (INP), 17489 Greifswald, Germany
[2]Institute of Physics, Faculty of Physics, Astronomy and Informatics, Nicolaus Copernicus University in Torun, 87–100 Toruń, Poland

[*] **E-mail:** ibrahim.sadiek@inp-greiswald.de

**Abstract:**

Non-thermal molecular plasmas play a crucial role in numerous industrial processes and hold significant potential for driving essential chemical transformations. Accurate information about the molecular composition of the plasmas and the distribution of populations among quantum states is essential for understanding and optimizing plasma processes. Here, we apply a mid-infrared frequency comb-based Fourier transform spectrometer to measure high-resolution spectra of plasmas containing hydrogen, nitrogen, and a carbon source in the 2800 – 3400 cm$^{-1}$ range. The spectrally broadband and high-resolution capabilities of this technique enable quantum-state-resolved spectroscopy of multiple plasma-generated species simultaneously, including $CH_4$, $C_2H_2$, $C_2H_6$, $NH_3$, and HCN, providing detailed information beyond the limitations of current methods. Using a line-by-line fitting approach, we analyzed 548 resolved transitions across five vibrational bands of plasma-generated HCN. The results indicate a significant non-thermal distribution of the populations among the quantum states, with distinct temperatures observed for lower and higher rotational quantum numbers, with a temperature difference of about 62 K. Broadband state-resolved-spectroscopy via comb-based methods provides unprecedented fundamental insights into the non-thermal nature of molecular plasmas – a detailed picture that has never been accomplished before for such complex non-thermal environment.

## 1. Introduction

Non-thermal molecular plasmas are increasingly used both in basic research and technology, playing essential roles in materials and surface processing, addressing environmental challenges, advancing plasma medicine, and facilitating tailored production in chemical synthesis [1-5]. Despite the continuously growing number of technological applications of molecular plasmas, questions regarding their non-thermal characteristics, in particular, the distribution of the population and energy among the quantum states of molecular species in plasmas, and the chemical reactivities of the plasma, remain under-investigated.

Reactive molecular plasmas, involving nitrogen ($N_2$), hydrogen ($H_2$), and a carbon source such as methane ($CH_4$) or carbon dioxide ($CO_2$) gases, or a screen of carbon-fiber-reinforced carbon (CFC), are used in several industrial processes, with plasma nitrocarburizing (PNC) as a prime example. With PNC the hardness and fatigue strength of metals can be enhanced through the diffusion of reactive nitrogen and carbon containing species into their surfaces. The demand for modified components via PNC processes is particularly high in sectors such as automotive, and aerospace industries [6,7]. Here, PNC serves as a model system for non-thermal low-pressure plasmas, where the plasma-induced chemistry differs fundamentally from the chemistry involved in conventional thermal processes. As a prerequisite to understand this chemistry and to subsequently model, optimize and control PNC processes, precise determination of non-thermal distributions of populations among the quantum states of plasma-generated species is essential.

Absorption spectroscopy has proven useful for providing absolute molecular densities and temperatures of species within the plasma [8-13]. Conventional laser absorption spectroscopy using narrow bandwidth continuous-wave (cw) lasers has been used for quantitative detection of stable and transient species in plasmas [8,9], particularly in the mid-infrared (mid-IR) region, where molecules exhibit strong absorptions due to their fundamental vibrations. However, accurate quantification of molecular densities, to some extent, and temperatures, to a large extent, of multiple species in plasmas remain challenging due to both the need for time-consuming tuning itself, as well as limited spectral tuning range of commonly used cw lasers.

Broadband detection techniques, such as Fourier-transform infrared (FT-IR) spectroscopy using incoherent light sources and optical dispersion spectroscopy using gratings, allow for the simultaneous detection of multiple species. However, the spectral resolution of conventional FT-IR spectrometers is limited by the scanning range of the optical path difference (OPD). For plasma-assisted applications such as PNC processes, operated at a typical pressure of a few mbar, the absorption profiles are expected to have a full width at half maximum (FWHM) of ~ 300 MHz (or 0.01 cm$^{-1}$). Consequently, a large OPD in the order of several meters will be required to spectrally resolve such narrow absorption profiles, making the resulting spectrometer bulky and challenging to align. Similarly, conventional grating-based dispersion spectrometers have limited spectral resolution determined by the resolving power of the gratings, typically not better than a few GHz. The limitations of restricted spectral coverage of absorption spectrometers based on cw lasers, and the restricted spectral resolution of conventional FT-IR and grating-based spectrometers can be overcome by employing an optical frequency comb as a light source.

The development of optical frequency combs has revolutionized the field of molecular spectroscopy. While most applications of direct frequency comb spectroscopy in the literature were focused on measurements under thermal equilibrium conditions [14-24], there have been few instances where frequency combs were applied to molecular plasmas: (i) A mid-IR optical frequency comb combined with a Fourier transform spectrometer has been used in a proof-of-feasibility experiment for down-stream analysis of the effluent from a dielectric barrier discharge [25], albeit with no state-resolved information from the measured spectra. (ii) Dual-comb spectroscopy has been used for time-resolved measurements of methane and ethane in an electric discharge in the mid-IR region around 3.3 μm [26], albeit with a relatively low spectral resolution of 6 GHz. However, the number of demonstrations of frequency comb spectroscopy for practical-oriented investigations – e.g., spectroscopy of reactive PNC processes – is very limited and mostly lacking quantum-state knowledge on the plasma-generated species. Very recently we have applied quantum cascade laser dual comb spectroscopy to investigate ammonia formation built in non-thermal $N_2$–$H_2$ plasmas [27]. Dual comb spectroscopy has also been applied to detect small traces of atomic Rb and K near 780 nm [28] as well as Fe near 533 nm [29] generated by laser-induced breakdown spectroscopy.

In this work, we utilize mid-IR frequency comb-based Fourier transform spectroscopy to measure high-resolution spectra of predominant molecular species produced in a low-pressure plasma containing $N_2$, $H_2$, and in direct contact with a screen of CFC, as a source of carbon. Several absorption bands of plasma-generated $CH_4$, $C_2H_2$, $C_2H_6$, $NH_3$, and HCN are measured simultaneously. A detailed line-by-line spectral analysis of resolved rovibrational transitions in five vibrational bands of HCN is performed using a Voigt line shape function. This enabled us to precisely determine the population and energy (temperature) distribution along the molecular vibrational ladder of HCN – a key molecular species that is believed to be linked to certain surface properties in PNC processes [30].

## 2. Experimental setup

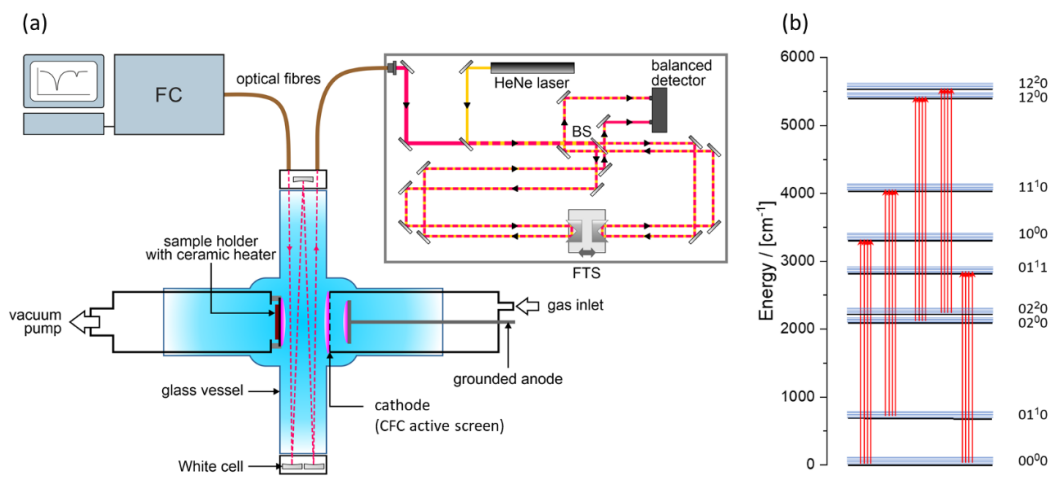

**Figure 1.** (a) Schematic of the frequency comb (FC)-based FTS setup coupled to a plasma test reactor. The test reactor has a multipass White cell attached to it. Optical fibers are used to couple the comb into the cell and from the cell to the FTS. The FTS includes a balanced detector, a frequency-stabilized Helium Neon (HeNe) laser for OPD calibration, and a beam splitter (BS) that works for both the FC and the HeNe laser. (b) Schematic of the molecular vibrational ladder of HCN (horizontal lines), along with the probed transitions (vertical arrows). The splitting of the $12^l0 \leftarrow 02^l0$ states with of $l = 0$ and 2 is depicted by two separate transitions in the ladder.

Figure 1(a) shows a schematic of the experimental setup, comprising a mid-IR frequency comb (Menlo Systems FC1500-250), a plasma nitrocarburizing monitoring reactor and an FTS. The mid-IR output, spanning from 2800 cm$^{-1}$ to 3400 cm$^{-1}$, was produced through difference frequency generation in a periodically poled lithium niobate crystal between spectrally shifted copies of a high-power Er-fiber comb, around 1550 nm, with a repetition frequency, $f_{rep}$, of 250 MHz. Consequently, the comb was inherently free from the carrier envelope offset frequency, $f_{ceo}$. The $f_{rep}$ was locked to the output of a tunable direct digital synthesizer referenced to a GPS-referenced Rb oscillator. The laser beam was coupled into a single-mode ZrF$_4$-BaF$_2$-LaF$_3$-AlF$_3$-NaF (ZBLAN) fiber (Thorlabs), and aligned to the multipass optics mounted to the reactor. The multipass cell (neoplas control) follows a so-called White arrangement capable of accommodating up to 32 passes. For the measurements reported here, 4 passes were used, resulting in an effective path length of 190 cm. The plasma was generated in a pulsed DC discharge at 1 kHz with a duty cycle of 60 %, operated with N$_2$ and H$_2$ gases. The spectroscopic investigations were carried on in a specially designed and downscaled plasma nitrocarburizing monitoring reactor, PLANIMOR, based on an industrial scale reactor, incorporating an active screen of CFC surrounding the workload, and allowing the gas flow to pass from the plasma region to the treated load, and hence provide the reactive species necessary for the hardening process [31]. Detailed information about the plasma reactor can be found elsewhere [32,33].

As shown schematically in Figure 1(a), the comb beam, after passing through the multipass cell, was coupled to the FTS via another single mode ZBLAN fiber. We used a *home-built* fast-scanning FTS with two retro-reflectors mounted back-to-back on a precision moving stage. The scan speed was set to 0.12 m s$^{-1}$, and the maximum OPD was 1.3 m, corresponding to a nominal resolution of 230 MHz (or 0.0077 cm$^{-1}$). For OPD calibration, we used a frequency-stabilized Helium Neon (HeNe) laser ($\lambda$ = 632.99 nm). Two out-of-phase comb interferograms were detected at the two outputs of the FTS using an auto-balancing HgCdTe detector (VIGO System, PVI-4TE-6) to minimize intensity noise [34]. The comb and HeNe laser interferograms were acquired using a digitizer (National Instruments, PCI-5922) and a LabVIEW$^{TM}$ program.

For high spectral resolution measurements, the sub-nominal resolution data analysis approach was used, as first developed [35,36], and recently used for measurements of complex spectra of heavy halogenated molecules [37,38], and for sub-Doppler measurements with a frequency comb as a probe[39], In the sub-nominal approach, the intensities of the comb modes are sampled very accurately by matching the nominal resolution of the FTS to the $f_{rep}$ of the comb, enabling absolute calibration of the frequency scale and substantially reducing instrumental line shape effects. At each measurement condition, the comb modes were stepped across the absorption spectrum by tuning the $f_{rep}$ of the comb. Specifically, we acquired spectra at four different $f_{rep}$ values separated by 173.75 Hz, corresponding to a 62 MHz step in the optical domain. For each $f_{rep}$ setting, a total of 50 interferograms were recorded, resulting in a total acquisition time of 34 min. Subsequently, the spectra recorded at the different $f_{\text{rep}}$ steps were normalized to a background spectrum recorded prior to switching the plasma on. After normalization, we calculated the corresponding absorption spectra using the Lambert-Beer law, and fitted a baseline consisting of a 4$^{\text{th}}$ order polynomial and a number of slowly varying sine terms to each spectrum while masking out the absorption lines. The baseline-corrected spectra were then interleaved to generate a final spectrum with a sampling point spacing of 62 MHz.

In our analysis we focus on HCN, as a key molecular species, due to its potential linkage with certain material properties in PNC processes itself [30]. Figure 1(b) shows a schematic energy level diagram of HCN, where the horizontal lines represent the vibrational levels of HCN, while the vertical arrows represent the probed transitions by the frequency comb. This figure displays five vibrational transitions involving: (i) the fundamental $v_1 \leftarrow v_0$ band, corresponding to the $10^00 - 00^00$ transition, (ii) the $v_1+v_2 \leftarrow v_2$ hot band, corresponding to the $11^10 - 01^10$ transition, (iii) the $v_1+2v_2 \leftarrow 2v_2$ hot band, corresponding to the $12^00 - 02^00$ transition, (iv) the $v_1+2v_2 \leftarrow 2v_2$ hot band, corresponding to the $12^20 - 02^20$ transition, and (v) the $v_2+v_3 \leftarrow v_0$ combination band, corresponding to the $01^11 - 00^00$ transition. It is worth noting that in Figure 1(b), the energy level diagram depicts separate states for the $12^l0 - 02^l0$ transitions with the quantum number of angular momentum $l = 0$ and 2.

## 3. Results and discussion
### 3.1. High-resolution spectroscopy of molecular

Figure 2(a) shows the measured high-resolution spectrum (black) of plasma-generated molecules in the 2800 – 3400 cm$^{-1}$ range along with models (inverted for clarity) based on the HITRAN [40] database and a Voigt lineshape function. In this experiment, the plasma was generated with N$_2$ and H$_2$ mass flow rates of 10 cm$^3$ min$^{-1}$ at standard conditions, a process pressure of 3 mbar, and a plasma power of 26 W. An active screen of CFC was used as a source of carbon in the PNC process under investigation; however, in more conventional applications of PNC, CH$_4$ or CO$_2$ is used as the source of carbon. For the models, we first assumed thermal population at a common $T = 370$ K, obtained from an average value of Doppler widths of resolved rovibrational transitions of HCN. The line intensities were also taken from the HITRAN [40], which assumes thermally equilibrated populations. The purpose of this exercise was to identify the molecular composition of the plasma before commencing a detailed spectral analysis of rovibrational state populations.

As shown in Figure 2(a), the spectra of five stable molecular species, namely CH$_4$, HCN, NH$_3$, C$_2$H$_2$, and C$_2$H$_6$ could be identified by comparison with the models. Although challenging to observe the absorption lines of C$_2$H$_6$ this low plasma power, the absorption becomes clearly observable at high plasma power. For HCN, there are two major rovibrational features dominated by the H$^{12}$C$^{14}$N isotopologue absorption. One feature appears around 3310 cm$^{-1}$, corresponding to the CH stretch vibration of the fundamental $v_1 \leftarrow v_0$ transition, as well as the $v_2+v_1 \leftarrow v_2$, and $2\nu_2+\nu_1 \leftarrow 2\nu_2$ hot band transitions. It is worth noting that the strong rovibrational transitions with low $J$ values ($J$ being the rotational quantum number) of the fundamental $\nu_1 \leftarrow \nu_0$ band were saturated, even at the lowest plasma power of 26 W investigated in this study. Another rovibrational feature is observed around 2810 cm$^{-1}$, corresponding to the simultaneous bending vibration and the CN stretching vibration, i.e., the $\nu_2+\nu_3 \leftarrow \nu_0$ combination band. Only the Q and R branches of the combination band are displayed in this range.

Zoomed-in spectral windows at different parts of the spectrum are presented in Figure 2(b) for the combination $\nu_2+\nu_4$ band of HCN around 2836.7 cm$^{-1}$, in Figure 2(c) for the Q-branch of the $\nu_3$ band of CH$_4$ around 3017.5 cm$^{-1}$, and in Figure 2(d) for the fundamental $\nu_1$ spectral region of H$^{12}$C$^{14}$N around 3331.5 cm$^{-1}$, where NH$_3$, C$_2$H$_2$, and H$^{13}$C$^{14}$N absorptions are also observable. The absorptions of the minor isotopologues of H$^{13}$C$^{14}$N and H$^{12}$C$^{15}$N become pronounced at higher plasma powers. A mismatch between the measured spectrum and the model was observed at several parts of the spectrum, e.g., see Figure 2(b), which clearly indicates a deviation from the assumed thermal population. This is further analyzed in the next section. Overall, the measured spectra in Figure 2 demonstrate the broadband and the high-resolution capabilities of optical frequency comb Fourier transform spectroscopy, which are particularly valuable for plasma diagnostic applications to precisely investigate the non-thermal population of quantum states in plasmas.

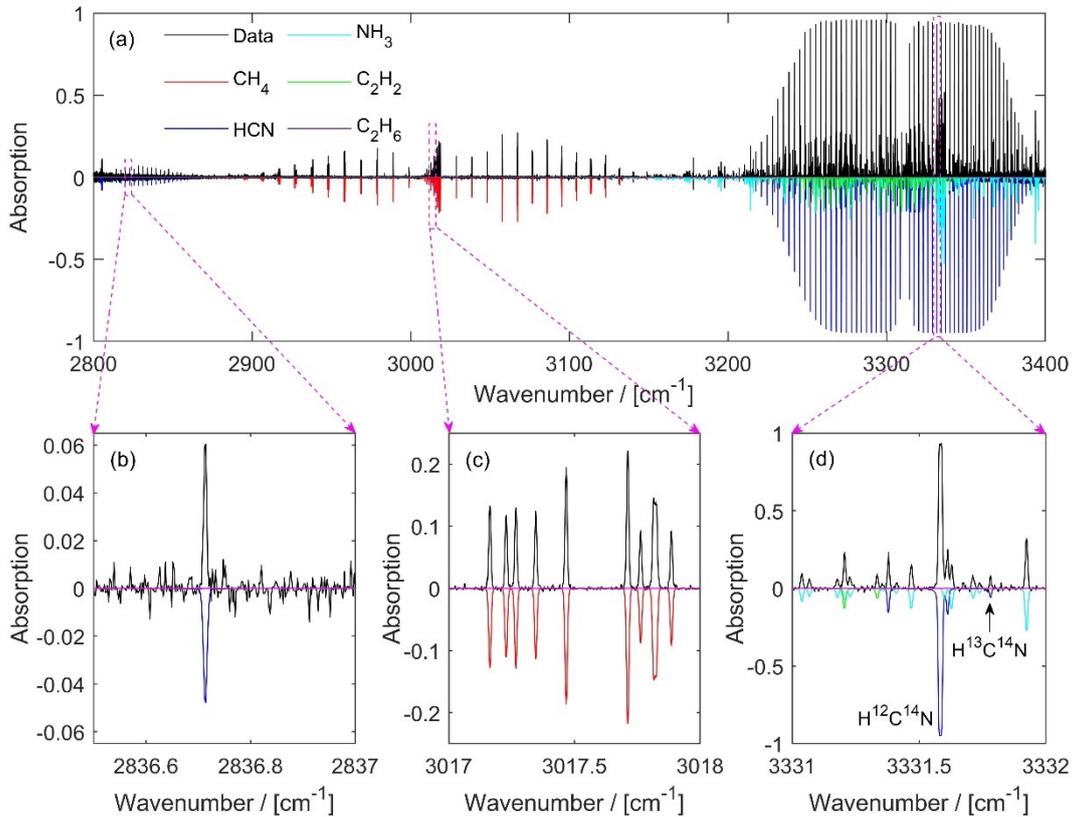

**Figure 2.** (a) High-resolution spectrum measured (black) in a PNC rector under pulsed DC discharge conditions of 1 kHz and a duty cycle of 60 %, process pressure of 3 mbar, and a plasma power of 26 W. The fitting models

(inverted for clarity) are based on the HITRAN [40] database and a Voigt line shape model assuming a single temperature of 370 K for all molecular species. (b) – (d) are zoomed-in portions of the spectrum, highlighting the state-resolved absorptions achieved for multiple molecular species in the plasma.

*3.2. Non-thermal population of rovibrational levels of HCN*

As already outlined earlier, HCN is a key molecule in all plasmas including $N_2$, $H_2$, and carbon sources, and it is believed to be linked to certain material properties in PNC processes [30]. Therefore, we put our focus on investigating the non-thermal distribution of populations among its quantum states, as a model system. We performed a line-by-line analysis of the resolved absorption profiles, corresponding to rovibrational transitions from state-*i* to state-*j*, using a Voigt line shape model with the fit parameters including the Doppler width and line center. As selection criteria for the absorption profiles used in the line-by-line analysis, a minimum absorption threshold was set at a signal-to-noise ratio (SNR) of 10, corresponding to an absorption of ~ 0.08, and a maximum absorption threshold of ~ 0.75 was set to avoid saturated absorption profiles. A total of 548 rovibrational transitions were analyzed for the five vibrational bands of HCN. Values of translational temperature, $T_{trans}$, were evaluated from the Doppler FWHM of the absorption profiles, while the population density in a lower level, $n_i$ was determined from the integral of the absorption profile normalized to the integrated absorption cross-section, $\sigma_{int}$. The value of $\sigma_{int}$ was evaluated based on the Einstein coefficient, $A_{ij}$, which unlike the spectral line intensity listed in HITRAN [40], is independent of whether local thermal equilibrium is assumed or not. By plotting $n_i$ values, each normalized to the statistical weight of the lower state, $g_i$, in a logarithmic scale as a function of the lower state energy, the so-called Boltzmann plots are obtained. The rotational temperatures ($T_{rot}$) are then determined from the slopes in these plots.

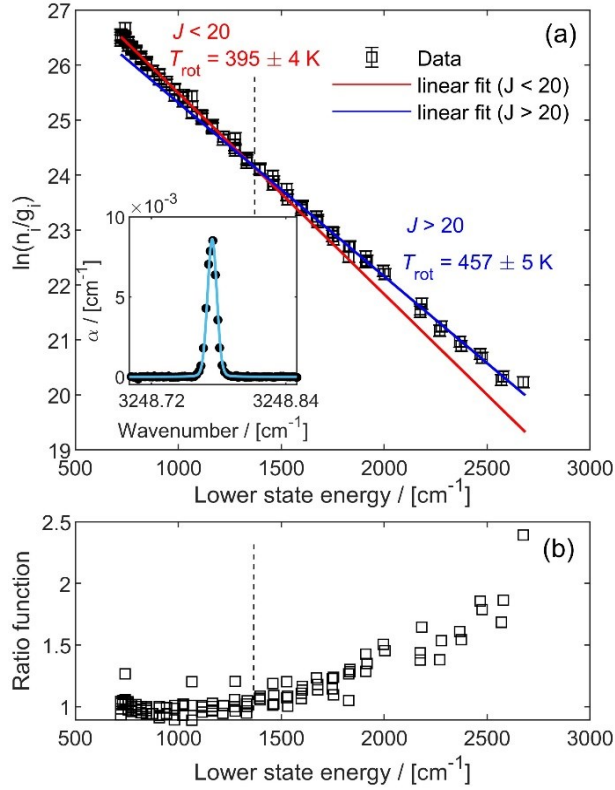

**Figure 3.** (a) Boltzmann plot for the $01^10$ hot band vibration of $H^{12}C^{14}N$ at a plasma power of 130 W. The red and blue lines are linear fits to the evaluated $n_i$ in rovibrational levels with $J < 20$ and $J > 20$, respectively. The inset shows an example of a measured absorption profile (black, dots) with the corresponding Voigt profile fit (blue). (b) A ratio function, defined as the ratio between all measured populations (squares in panel a) relative to the distribution of population in $J < 20$ levels (red linear fit in panel a).

In Figure 3(a) the evaluated Boltzmann plot (black squares) from rovibrational transitions of the $11^10 \leftarrow 01^10$ hot band at a plasma power of 130 W is shown. The inset of Figure 3(a) shows an example of a fitted profile, where the integral is proportional to $n_i$. Notably, Figure 3(a) exhibits two distinct rotational temperatures, as evidenced by the linear fits applied to levels with $J < 20$ (indicated by the red solid line) and levels with $J > 20$ (indicated by the blue solid line), indicating non-thermal population of rotational levels of the $01^10$ state of HCN. Figure 3(b) illustrates a ratio function describing the evolution of population in all rotational levels relative to the distribution of population in low $J$ levels ($J < 20$). A value of greater than 1 indicates overpopulation relative to lower $J$ levels. As can be seen in this figure, up to lower state energy of ~ 1400 cm$^{-1}$ ($J = 20$), the ratio is equal 1, while for rotational levels with $J > 20$, an overpopulation is observed, which can be described by another rotational temperature.

The fitting of two slopes to the determined rovibrational population densities indicates that a 'single temperature' describing the distribution of populated quantum states in plasmas is not valid. Clearly, such detailed non-thermal picture could only be obtained through broadband and highly precise absorption profile measurements provided by frequency comb spectroscopy.

We extended the detailed line-by-line analysis for spectra of HCN measured at different plasma powers of 45 W, and 93 W. In Table 1, the evaluated $T_{rot}$ and $T_{trans}$ from the $11^10 \leftarrow 01^10$ hot band rovibrational transitions at four different plasma powers are given. Two distinct $T_{rot}$ values for the $01^10$ rovibrational state were observed at a plasma power of 93 W, as was the case at 130 W [Figure 3(a)]. However, the difference between the two evaluated $T_{rot}$ values decreased from 62 K at 130 W to 55 K at 93 W. At a further lower plasma power of 45 W, the evaluated $T_{rot}$ values from the different parts of the Boltzmann plot show similar values. As one may expect, both $T_{rot}$ and $T_{trans}$ increase with increasing plasma power [Table 1]. It is worth to note that $T_{rot}$ values for both $J < 20$ and $J > 20$ levels are lower than the $T_{trans}$ at each plasma power. This might be attributed to the efficient energy relaxation mechanisms in HCN, involving the very fast vibrational-vibrational/rotational (V - V/R) energy transfer from rovibrationally excited states to the lowest vibrational mode (the $\nu_2$ mode) [41]. Consequently, the vibrational and rotational energies are dissipated as heat in a form of translational motion (i.e., via V – T and R – T processes), resulting in higher $T_{trans}$ values for the ground state compared to $T_{rot}$. This observation aligns with the large dipole moment of HCN (2.980 Debye) [42], and with its ability to form hydrogen bonds, which makes it an efficient energy relaxation collider similar to molecules such as water and $NH_3$ [43].

In Table 2, we present a summary of temperatures obtained at three plasma power: 130 W, 93 W, and 45 W for all analyzed vibrational states of HCN. For some vibrational states, only one rotational temperature could be evaluated because of either low SNR (particularly for $J > 20$ levels), or signal saturation (particularly for $J < 20$ levels and at high plasma powers). Nevertheless, the following key conclusions can be drawn from Table 2, specifically for vibrational states involving one vibrational quantum (the first three rows in of Table 2): (i) At a given plasma power, $T_{trans}$ values obtained from Doppler widths of rovibrational transitions are similar for all vibrational states. (iii) $T_{rot}$ increases with the lower energy of the vibrational states (i.e., $T_{rot}(2\nu_2) > T_{rot}(\nu_2) > T_{rot}(\nu_0)$) ). Furthermore, at a given plasma power, $T_{trans}$ values connected to transitions involving two vibrational quanta (i.e., the $01^11 \leftarrow 00^00$ transition: the last row in Table 2) are lower than $T_{trans}$ values connected to transitions involving one vibrational quantum (i.e., the $10^00 \leftarrow 00^00$, $11^10 \leftarrow 01^10$, and $02^l0 \leftarrow 12^l0$ transitions). This observation may suggest the presence of independent energy relaxation channels of HCN. This implies the existence of competing V – T pathways instead of a single V – T process through the lowest vibrational mode, as is the case for the majority of polyatomic molecules under thermal conditions [41]. Under such conditions, the fast V – V processes distribute the vibrational quanta among the states, then the vibrational quanta are being dissipated through the lowest vibrational mode via V – T processes. It may also indicate the presence of different formation pathways, such as the formation of HCN in different rovibrationally excited states, in the non-thermal plasmas.

**Table 1.** Values of rotational and translational temperatures for the $01^10$ vibrational state of HCN at different plasma powers. The uncertainties in $T_{rot}$ represent $1\sigma$ confidence level of the slope parameter in the linear fits, while in $T_{trans}$, they represent a $1\sigma$ of Doppler width parameter fits to line profiles.

| Plasma power [W] | $T_{rot}$ [K] | | $T_{trans}$ [K] |
|---|---|---|---|
| | $J > 20$ | $J < 20$ | |
| 130 | 395 ± 4 | 457 ± 5 | 478 ± 21 |
| 93 | 369 ± 3 | 424 ± 11 | 439 ± 27 |
| 45 | 347 ± 3 | 349 ± 11 | 390 ± 13 |
| 26 | 313 ± 4 | | 363 ± 22 |

**Table 2.** Values of rotational and translational temperatures involved in five vibrational transitions of HCN at different plasma powers. The uncertainties in $T_{rot}$ represent 1σ confidence level of the slope parameter in the linear fits, while in $T_{trans}$, they represent 1σ of Doppler width parameter fits to line profiles.

| Transition | Plasma power = 130 W | | | Plasma power = 93 W | | | Plasma power = 45 W | | |
|---|---|---|---|---|---|---|---|---|---|
| | $T_{rot}$ [K] | | $T_{trans}$ [K] | $T_{rot}$ [K] | | $T_{trans}$ [K] | $T_{rot}$ [K] | | $T_{trans}$ [K] |
| | $J < 20$ | $J > 20$ | | $J < 20$ | $J > 20$ | | $J < 20$ | $J > 20$ | |
| $10^00 \leftarrow 00^00$ | | 436 ± 6 | 487 ± 20 | | 393 ± 6 | 440 ± 24 | | 332 ± 6 | 390 ± 8 |
| $11^10 \leftarrow 01^10$ | 395 ± 4 | 457 ± 5 | 478 ± 21 | 369 ± 3 | 424 ± 11 | 439 ± 27 | 347 ± 13 | 349 ± 11 | 390 ± 13 |
| $12^l0 \leftarrow 02^l0$ a) | 410 ± 13 | | 478 ± 41 | | | 437 ± 40 | | | |
| $01^11 \leftarrow 00^00$ | 331 ± 4 | | 432 ± 22 | 315 ± 5 | | 407 ± 23 | 313 ± 5 | | 398 ± 24 |

a): Splitting of the $12^l0 \leftarrow 02^l0$ states with different values of $l$ are not shown; Values presented are average over $l$=0 and 2 states.

To further illustrate the observed non-thermal population of rovibrational transitions of HCN generated in PNC processes, we utilized the determined temperatures from Table 2 in a broadband spectral model based on the HITRAN database [40] and a Voigt line shape function. Here, the line intensities were evaluated based on the Einstein-$A_{ij}$ coefficient, which is unlike the line strengths listed in HITRAN [40], is independent whether a thermal equilibrium exist or not. Figure 4(a) shows the measured spectrum (black) at a plasma power of 93 W in the range of 3200 – 3400 cm$^{-1}$, where the four $\nu_1$ stretching vibrations of HCN are located, together with the model spectrum. Figure 4 (b) – (d) shows the residuals of the fit to the model under different scenarios: (b) when only one temperature is included in the model, i.e., assuming thermally populated rovibrational transitions with $T_{rot} = T_{trans}$, (c) when one $T_{trans}$ is shared among all the transitions, and one $T_{rot}$ is assigned to each vibrational band, and (d) when one $T_{trans}$ is shared among all the transitions and two $T_{rot}$ are included. It should be noted that the two $T_{rot}$ values were only included for the $10^00 \leftarrow 00^00$, an $11^10 \leftarrow 01^10$ vibrational band, as the other vibrational bands have relatively weak absorption signals. Additionally, the absorption contributions from NH$_3$ and C$_2$H$_2$ were considered in all three scenarios with one $T_{trans}$ and one $T_{rot}$. A detailed analysis of the rovibrational spectrum of these molecules (i.e., whether one $T_{rot}$ is sufficient to model the spectra of each vibrational band or more than one $T_{rot}$) will be further conducted in future work.

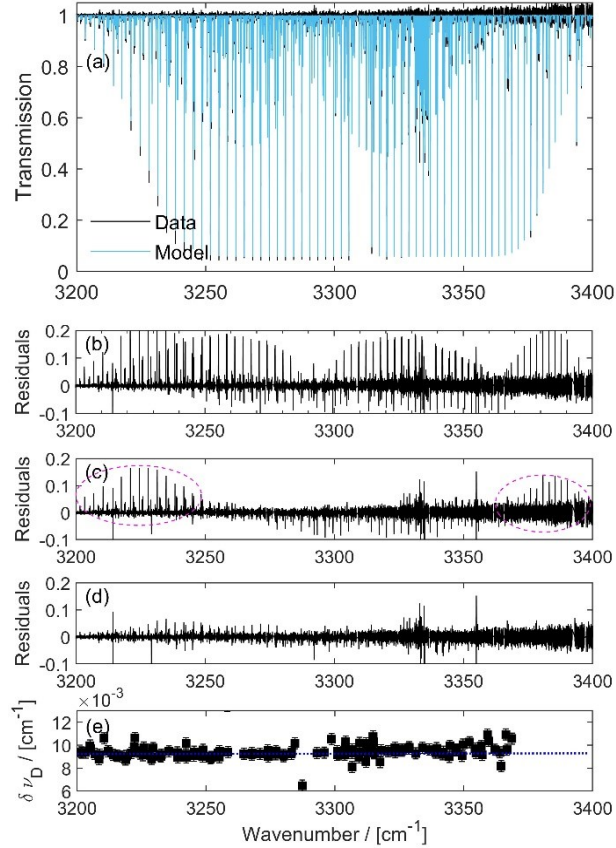

**Figure 4.** (a) High-resolution spectrum measured (black) at a plasma power of 93 W, together with simulations (blue) utilizing the HITRAN database and a Voigt line shape function. The model accounts for different rotational temperatures for each vibrational band. (b) Residuals when only one temperature is used in the model ($T_{tran} = T_{rot}$) for all vibrations of HCN. (c) Residuals when one $T_{trans}$ is used for all the states and one $T_{rot}$ is used for each vibration. The ellipses highlight the regions of high $J$ values with large residuals. (d) Residuals when one $T_{trans}$ is used for all the states and two $T_{rot}$ are used for each vibration. (e) The Doppler FWHM, $\delta\nu_D$, values obtained from the line-by-line fitting of rovibrational transitions from the four bands of HCN. The dotted line indicates the mean value.

In Figure 4(b), where only one temperature was used in the model for HCN, large residuals are observed in all the vibrational bands – indicating that the concept of a single temperature describing the distribution of all molecular states is not valid. It also shows the need to include different $T_{rot}$ for the different vibrational bands in the model, as illustrated in Figure 4(c), where one $T_{rot}$ is assigned to each vibration. The residuals are significantly reduced; however, some residuals still remain at higher $J$ values for both the $10^00 \leftarrow 00^00$ and $11^10 \leftarrow 01^10$ transitions. By including two $T_{rot}$ for each of the $10^00 \leftarrow 00^00$ and $11^10 \leftarrow 01^10$ transitions [Figure 4(d)], the residuals are further reduced, aligning with the earlier line-by-line analysis results presented. Overall, these results demonstrate the need for multiple rotational temperatures in the model to accurately describe the observed non-thermal population of rovibrational states of HCN. Figure 5 depicts a zoomed-in window of Figure 4(a) together with the fit residuals for the scenario when two rotational temperatures are included in the simulations of the $10^00 \leftarrow 00^00$ and $11^10 \leftarrow 01^10$ rovibrational transitions.

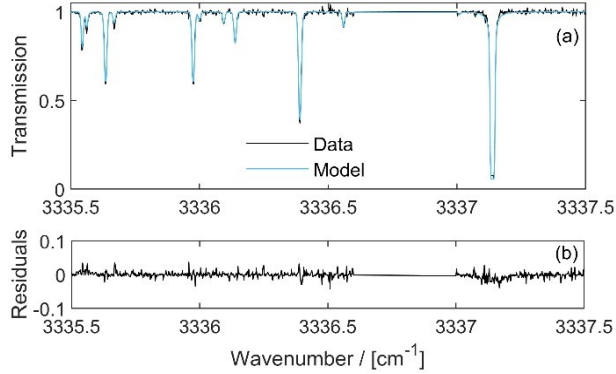

**Figure 5.** (a) Zoomed-in window of spectrum shown in Figure 4(a) showing the measured spectra (black) together with overall model (blue) including two rotational temperatures for each of the $\nu_0$ and $\nu_2$ states. (b) Residual of the fit. The masked region around 3336.6 cm$^{-1}$ corresponds to a location of a water line in this range.

Finally, considering that absorption spectroscopy is a line-of-sight method with an integrating character, local heating in the active plasma region may be expected to induce a temperature gradient, in particular at high plasma powers. If this is the case, the measured absorption profiles, particularly their Doppler width, will respond differently to such a gradient based on whether the probed transitions are from a hot band or a fundamental band, or whether they are for high $J$ or low $J$ transitions [44-46]. For instance, higher $J$ and hot band transitions are predicted to include more contribution from the hotter region in the reactor than the low $J$ transitions of the fundamental band, as expected from the dependence of their spectral line intensity on temperature [40]. In our plasma test reactor, where we have not implemented any additional heating and the used plasma powers were relatively low, we anticipate a conservative temperature gradient of ~ 50 K. The uncertainty in the determined temperatures due to this gradient is within the 5% average uncertainty of the Doppler widths evaluated from the line-by-line fitting. Furthermore, in Figure 4(e), the Doppler widths (FWHM), $\delta\nu_D$ obtained from the line-by-line fitting of resolved rovibrational transitions at a plasma power of 93 W are shown. As can be seen in this figure, the values of $\delta\nu_D$ are randomly scattered along a mean value (indicated by the blue dotted line) of ~0.0095 cm$^{-1}$, with no tendency to increase at higher $J$ values or for hot band transitions. This suggests a negligible contribution of the temperature gradient inside the plasma test reactor to the observed non-thermal population of rovibrational levels of HCN given the SNR of the measurements.

## 4. Conclusion

We have utilized the broad bandwidth and high spectral resolution capabilities of frequency comb Fourier transform spectroscopy for investigating non-thermal characteristics of molecular plasmas. Through quantum-state-resolved spectroscopy, we have precisely determined the distribution of populations among the quantum states of plasma-generated species. We used reactive plasmas of $N_2$, $H_2$ and screen of CFC (as a carbon source) – a technology known as plasma nitrocarburizing (PNC) – as a model system for low-pressure non-thermal molecular plasmas. Within the spectral range from 2800 cm$^{-1}$ to 3400 cm$^{-1}$, we measured several rovibrational bands of CH$_4$, C$_2$H$_2$, C$_2$H$_6$, NH$_3$, and HCN generated in the plasma. Using a line-by-line fitting approach with a Voigt line shape function, we analyzed a total of 548 resolved rovibrational transitions across five vibrational bands of HCN. The precise determination of the rotational and translational temperatures for these states revealed a non-thermal population distribution within the vibrational levels. Notably, we observed two distinct rotational temperatures for lower $J$ and higher $J$ rotational transitions of individual vibrational states, with a difference of up to 62 K. This temperature difference decreased with the plasma power. The spectrally broadband and high-resolution frequency comb system enabled us to probe several rovibrational transitions along the vibrational ladder of HCN. This provides detailed information about the energy (temperature) and population distributions within the quantum states of the molecule beyond the limitations of currently used methods. Such detailed information paves the way to use quantum state populations as quantitative indicators to tailor the chemical synthesis in plasma processes and guide them to provide the desired outcome.

**Acknowledgement**

The authors acknowledge funding for this project from the Deutsche Forschungsgemeinschaft (DFG, German Research Foundation) - project No SA 4483/1–1. A. N. was supported by National Science Centre, Poland project no. 2019/35/D/ST2/04114. We acknowledge valuable comments from A. Foltynowicz and A. J. Fleisher, and technical support from U. Macherius and F. Weichbrodt.

**Conflict of Interest**

The authors declare no conflict of interest.

**Data availability**

The data that support the findings of this study are available from the corresponding author upon reasonable request.

[1] Angus J C and Hayman C C 1988 Metastable Growth of Diamond and Diamondlike Phases *Science* **241**, 913–921.

[2] Weltmann K D, Brandenburg R, von Woedtke T, Ehlbeck J, Foest R, Stieber M and Kindel E 2008 Antimicrobial treatment of heat sensitive products by miniaturized atmospheric pressure plasma jets (APPJs) *J. Phys. D: Appl. Phys.* **41**, 194008.

[3] Brandenburg R, Bogaerts A, Bongers W, Fridman A, Fridman G, Locke B R, Miller V, Reuter S, Schiorlin M, Verreycken T and Ostrikov K 2019 Metastable growth of diamond and diamond-like phases *Plasma Processes and Polym.* **16**, 1700238.

[4] Bekeschus S, von Woedtke T, Emmert S and Schmidt A 2021 Medical gas plasma-stimulated wound healing: Evidence and mechanisms *Redox Biol.* **46**, 102116.

[5] Adamovich I, Agarwal S, Ahedo E, Alves L L, Baalrud S, Babaeva N, Bogaerts A, Bourdon A, Bruggeman P J, Canal C, Choi E H, Coulombe S, Donkó Z, Graves D B, Hamaguchi S, Hegemann D, Hori M, Kim H H, Kroesen G M W, Kushner M J, Laricchiuta A, Li X, Magin T E, Thagard S M, Miller V, Murphy A B, Oehrlein G S, Puac N, Sankaran R M, Samukawa S, Shiratani M, Šimek M, Tarasenko N, Terashima K, Thomas Jr E, Trieschmann J, Tsikata S, Turner M M, van der Walt I J, van de Sanden M C M and von Woedtke T 2022 Plasma Roadmap: low temperature plasma science and technology *J. Phys. D: Appl. Phys.* **55**, 373001.

[6] Grün R and Günther H-J 1991 Plasma nitriding in industry—problems, new solutions and limits *Mater. Sci. Eng. A* **140**, 435–44.

[7] Michel H, Czerwiec T, Gantois M, Ablitzer D and Ricard A 1995 Progress in the analysis of the mechanisms of ion nitriding *Surf. Coat. Technol.* **72**, 103–111.

[8] Hempel F, Davies P B, Loffhagen D, Mechold L and Röpcke J 2003 Diagnostic studies of $H_2$–Ar–$N_2$ microwave plasmas containing methane or methanol using tunable infrared diode laser absorption spectroscopy *Plasma Sources Sci Technol.* **12**, S98.

[9] van Helden J H, van den Oever P J, Kessels W M M, van de Sanden M C M, Schram D C and Engeln R 2007 Production Mechanisms of NH and $NH_2$ Radicals in $N_2$−$H_2$ Plasmas," *J. Phys. Chem. A* **111**, 11460–11472.

[10] van Helden J H, Hancock G, Peverall R and Ritchie G A D 2011 A 3 μm difference frequency laser source for probing hydrocarbon plasmas *J. Phys. D: Appl. Phys.* **44**, 125202.

[11] Röpcke J, Davies P B, Lang N, Rousseau A and Welzel S 2012 Applications of quantum cascade lasers in plasma diagnostics: a review *J. Phys. D: Appl. Phys.* **45**, 423001.

[12] Puth A, Kusýn L, Pipa A V, Burlacov I, Dalke A, Hamann S, van Helden J. H, Biermann H and Röpcke J 2020 Spectroscopic study of plasma nitrocarburizing processes with an industrial-scale carbon active screen *Plasma Sources Sci Technol.* **29**, 035001.

[13] Dekkar D, Puth A, Bisceglia E, Moreira P W P, Pipa A V, Lombardi G, Röpcke J, van Helden J H and Bénédic F 2020 Effect of the admixture of $N_2$ to low pressure, low temperature $H_2$–$CH_4$–$CO_2$ microwave plasmas used for large area deposition of nanocrystalline diamond films *J. Phys. D: Appl. Phys.* **53**, 455204.

[14] Marian A, Stowe M C, Lawall J R, Felinto D and Ye J 2004 United time-frequency spectroscopy for dynamics and global structure *Science* **306**, 2063-2068.

[15] Gherman T, Kassi S, Campargue A and Romanini D 2004 Overtone spectroscopy in the blue region by cavity-enhanced absorption spectroscopy with a mode-locked femtosecond laser: application to acetylene *Chem. Phys. Lett.* **383**, 353-358.

[16] Tillman K A, Maier R R J, Reid D T and McNaghten E D 2005 Mid-infrared absorption spectroscopy of methane using a broadband femtosecond optical parametric oscillator based on aperiodically poled lithium niobate *J. Opt. A: Pure Appl. Opt.* **7**, S408.

[17] Thorpe M J, Moll K D, Jones R J, Safdi B and Ye J 2006 Broadband cavity ringdown spectroscopy for sensitive and rapid molecular detection *Science* **311**, 1595-1599.

[18] Diddams S A, Hollberg L and Mbele V 2007 Molecular fingerprinting with the resolved modes of a femtosecond laser frequency comb *Nature* **445**, 627–630.

[19] Adler F, Masłowski P, Foltynowicz A, Cossel K C, Briles T C, Hartl I and Ye J 2010 Mid-infrared Fourier transform spectroscopy with a broadband frequency comb *Opt. Express* **18**, 21861–21872.

[20] Sadiek I, Mikkonen T, Vainio M, Toivonen J and Foltynowicz A 2018 Optical frequency comb photoacoustic spectroscopy *Phys. Chem. Chem. Phys.* **20**, 27849–27855.

[21] Muraviev A V, Smolski V O, Loparo Z E and Vodopyanov K L 2018 Massively parallel sensing of trace molecules and their isotopologues with broadband subharmonic mid-infrared frequency combs *Nat. Photonics* **12**, 209–214.

[22] Timmers H, Kowligy A, Lind A, Cruz F C, Nader N, Silfies M, Ycas G, Allison T K, Schunemann P G, Papp S B and Diddams S A 2018 Molecular fingerprinting with bright, broadband infrared frequency combs *Optica* **5**, 727–732.

[23] Fortier T and Baumann E 2019 20 years of developments in optical frequency comb technology and applications *Commun. Phys.* **2**, 153.

[24] Sadiek I, Hjältén A, Senna Vieira F, Lu C, Stuhr M and Foltynowicz A 2020 Line positions and intensities of the $\nu_4$ band of methyl iodide using mid-infrared optical frequency comb Fourier transform spectroscopy *J. Quant. Spectrosc. Radiat. Transfer* **255**, 107263.

[25] Golkowski M, Golkowski C, Leszczynski J, Plimpton S R, Maslowski P, Foltynowicz A, Ye J and McCollister B 2012 Hydrogen-peroxide-enhanced nonthermal plasma effluent for biomedical applications *IEEE Trans. Plasma Sci.* **40**, 1984–1991.

[26] Abbas M A, Pan Q, Mandon J, Cristescu S M, Harren F J M and Khodabakhsh A 2019 Time-resolved mid-infrared dual-comb spectroscopy *Sci. Rep.* **9**, 17247.

[27] Sadiek I, Fleisher A J, Hayden J, Huang X, Hugi A, Engeln R, Lang N, and van Helden J H 2023 Dual-comb spectroscopy of ammonia formation in non-thermal plasmas," arXiv:2312.05075.

[28] Bergevin J, Wu T-H, Yeak J, Brumfield B E, Harilal S S, Phillips M C and Jones R J 2018 Dual-comb spectroscopy of laser-induced plasmas *Nat. Commun.* **9**, 1273.

[29] Zhang Y, Lecaplain C, Weeks R R D, Yeak J, Harilal S S, Phillips M C and Jason Jones R 2019 Time-resolved dual-comb measurement of number density and temperature in a laser-induced plasma *Opt. Lett.* **44**, 3458–3461.

[30] Dalke A, Burlacov I, Hamann S, Puth A, Böcker J, Spies H-J, Röpcke J and Biermann H 2019 Solid carbon active screen plasma nitrocarburizing of AISI 316L stainless steel: Influence of $N_2$-$H_2$ gas composition on structure and properties of expanded austenite *Surf. Coat. Technol.* **357**, 1060–1068.

[31] Li C X, Georges J and Li X Y 2002 Active screen plasma nitriding of austenitic stainless steel *Surf. Eng.* **18**, 453–45.

[32] Hamann S, Burlacov I, Spies H-J, Biermann H and Röpcke J 2017 Spectroscopic investigations of plasma nitriding processes: A comparative study using steel and carbon as active screen materials *J. Appl. Phys.* **121**, 153301.

[33] Hamann S, Börner K, Burlacov I, Spies H-J, Strämke M, Strämke S and Röpcke J 2015 Plasma nitriding monitoring reactor: A model reactor for studying plasma nitriding processes using an active screen *Rev. Sci. Instrum.* **86**, 123503.

[34] Foltynowicz A, Ban T, Masłowski P, Adler F and Ye J 2011 Quantum-noise-limited optical frequency comb spectroscopy *Phys. Rev. Lett.* **107**, 233002.

[35] Maslowski P, Lee K F, Johansson A C, Khodabakhsh A, Kowzan G, Rutkowski L, Mills A A, Mohr C, Jiang J, Fermann M E and Foltynowicz A 2016 Surpassing the path-limited resolution of Fourier-transform spectrometry with frequency combs *Phys. Rev. A* **93**, 021802.

[36] Rutkowski L, Masłowski P, Johansson A C, Khodabakhsh A and Foltynowicz A 2018 Optical frequency comb Fourier transform spectroscopy with sub-nominal resolution and precision beyond the Voigt profile *J. Quant. Spectrosc. Radiat. Transfer* **204**, 63–73.

[37] Sadiek I, Hjältén A, Roberts F C, Lehman J H and Foltynowicz A 2023 Optical frequency comb-based measurements and the revisited assignment of high-resolution spectra of $CH_2Br_2$ in the 2960 to 3120 $cm^{-1}$ region *Phys. Chem. Chem. Phys.* **25**, 8743-8754.

[38] Hjältén A, Foltynowicz A and Sadiek I 2023 Line positions and intensities of the $\nu_1$ band of $^{12}CH_3I$ using mid-infrared optical frequency comb Fourier transform spectroscopy *J. Quant. Spectrosc. Radiat. Transfer* **306**, 108646.

[39] Foltynowicz A, Rutkowski L, Silander I, Johansson A C, Silva de Oliveira V, Axner O, Soboń G, Martynkien T, Mergo P and Lehmann K K 2021 Sub-Doppler double-resonance spectroscopy of methane using a frequency comb probe *Phys. Rev. Lett.* **126**, 063001.

[40] Gordon I E, Rothman L S, Hargreaves R J, Hashemi R, Karlovets E V, Skinner F M, Conway E K, Hill C, Kochanov R V, Tan Y, Wcisło P, Finenko A A, Nelson K, Bernath P F, Birk M, Boudon V, Campargue A, Chance K V, Coustenis A, Drouin B J, Flaud J M, Gamache R R, Hodges J T, Jacquemart D, Mlawer E J, Nikitin A V, Perevalov V I, Rotger M, Tennyson J, Toon G C, Tran H, Tyuterev V G, Adkins E M, Baker A, Barbe A, Canè E, Császár A G, Dudaryonok A, Egorov O, Fleisher A J, Fleurbaey H, Foltynowicz A, Furtenbacher T, Harrison J J, Hartmann J M, Horneman V M, Huang X, Karman T, Karns J, Kassi S, Kleiner I, Kofman V, Kwabia–Tchana F, Lavrentieva N N, Lee T J, Long D A, Lukashevskaya A A, Lyulin O M, Makhnev V Y, Matt W, Massie S T, Melosso M, Mikhailenko S N, Mondelain D, Müller H S P, Naumenko O V, Perrin A, Polyansky O L, Raddaoui E, Raston P L, Reed Z D, Rey M, Richard C, Tóbiás R, Sadiek I, Schwenke D W, Starikova E, Sung K, Tamassia F, Tashkun S A, Vander Auwera J, Vasilenko I A, Vigasin A A, Villanueva G L, Vispoel B, Wagner G, Yachmenev A and Yurchenko S N 2022 The HITRAN2020 molecular spectroscopic database *J. Quant. Spectrosc. Radiat. Transfer* **277**, 107949.

[41] Lambert J D 1972 Vibration-translation and vibration-rotation energy transfer in polyatomic molecules *J. Chem. Soc., Faraday Trans. 2* **68**, 364–373.

[42] DeLeon R L and Muenter J S 1984 The vibrational dipole moment function of HCN *J. Chem. Phys.* **80**, 3992-3999.

[43] Huang R, Wu J, Gong M-X., Saury A and Carrasquillo M E 1993 Direct observation of vibrational relaxation in highly excited HCN *Chem. Phys. Lett.* **216**, 108–114.

[44] Nave A S C, Baudrillart B, Hamann S, Bénédic F, Lombardi G, Gicquel A, van Helden J H and Röpcke J 2016 Spectroscopic study of low pressure, low temperature $H_2$-$CH_4$-$CO_2$ microwave plasmas used for large area deposition of

nanocrystalline diamond films. Part I: on temperature determination and energetic aspects *Plasma Sources Sci. Technol.* **25**, 065002.

[45] Pipa A V, Puth A, Böcker J, Jafarpour S M, Dalke A, Biermann H, Röpcke J and van Helden J H 2023 Laser absorption spectroscopy for plasma-assisted thermochemical treatment. Part I.: Applicability of the Beer–Lambert law and interpretation of spectroscopic data *Plasma Sources Sci. Technol.* **32**, 085011.

[46] Pipa A V, Puth A, Böcker J, Jafarpour S M, Dalke A, Biermann H, Röpcke J and van Helden J H 2023 Laser absorption spectroscopy for plasma-assisted thermochemical treatment. Part II.: Impact of the carbon and water contaminants on a low-pressure $N_2$–$H_2$ discharge *Plasma Sources Sci. Technol.* **32**, 085012.